\documentclass[a4, 11pt]{article}
\title{On  fermion   
grading symmetry} 
\date{}
\author{Hajime Moriya}
\usepackage{latexsym}
\usepackage{amssymb} 
\usepackage{amsmath} %
\usepackage{amsthm} %
\usepackage{amscd} %
\newcommand{\R} {{\mathbb{R}}}%
\newcommand{\Z}{{\mathbb{Z}}}%
\newcommand{\CC} {{\mathbb{C}}}%
\newcommand{\NN} {{\mathbb{N}}}%
\newcommand{\physical}{\Lambda}%
\newcommand{\physicalG}{{\physical}^G}%
\newcommand{\Znu}{\Z^{\nu}}%
\newcommand{\id}{{\mathbf{1} } }
\newcommand{\proofend}{{\hfill $\Box$}}
\newcommand{\cstar}{{\bf C}^{\ast}}%
\newcommand{\nonum}{\nonumber}%
\newcommand{\Imag}{{\bf{Im}}}%
\newcommand{\Real}{{\bf{Re}}}%
\newcommand{\vp}{\varphi}
\newcommand{\ome}{\omega}
\newcommand{\vpx}{\vp_{\xi}}
\newcommand{\del}{\delta}
\newcommand{\pot}{{\mit \Phi}}%
\newcommand{\pott}{ \tilde{\pot} }%
\newcommand{\altt}{ \tilde{\al}_t }%
\newcommand{\HIt}{ \tilde{H}(\I)}
\newcommand{\HJt}{ \tilde{H}(\J)}
\newcommand{\delt}{\tilde{\del}}
\newcommand{\al}{\alpha}%
\newcommand{\alt}{\alpha_t}%
\newcommand{\alp}{\al^{\prime}}%
\newcommand{\alc}{\al_c}%
\newcommand{\gammac}{\gamma_{c}}%
\newcommand{\I}{{\mathrm{I}}}%
\newcommand{\J}{{\mathrm{J}}}%
\newcommand{\K}{{\mathrm{K}}}%
\newcommand{\Ic}{\I_{\rm{c}}}%
\newcommand{\oric}{\{0\}_{\rm{c}}}%
\newcommand{\Fin}{{\mathfrak F}}%
\newcommand{\Finf}{\Fin_{ {\rm{loc}} }}%
\newcommand{\Hil}{{\cal H}}%
\newcommand{\Kil}{{\cal K}}%
\newcommand{\UIc}{U_{\Ic}}
\newcommand{\omei}{\ome_{i}}
\newcommand{\PB}{{\cal P}} %
\newcommand{\vareps}{\varepsilon}%
\newcommand{\psitil}{\tilde{\psi}}
\newcommand{\omef}{\ome_{1}}
\newcommand{\omes}{\ome_{2}}
\newcommand{\omet}{{\ome\Theta}}
\newcommand{\vpHI}{\vp^{\beta\HI} }%
\newcommand{\qp}{q_{+} }%
\newcommand{\qm}{q_{-} }%
\newcommand{\pf}{p_{f} }%
\newcommand{\pfm}{p_{-f} }%
\newcommand{\vpHIp}{\vpHI_{<p>} }%
\newcommand{\vpHIpf}{\vpHI_{<\pf>} }%
\newcommand{\vpHIpfm}{\vpHI_{<\pfm>} }%
\newcommand{\vpHIIc}{{\vpHI}_{\Ic}}
\newcommand{\vpIc}{\vp_{\Ic}}
\newcommand{\aicr}{a_i^{\ast}}%
\newcommand{\ai}{a_i}%
\newcommand{\piome}{\pi_{\ome}}
\newcommand{\pipsitil}{\pi_{\psitil}}
\newcommand{\pivp}{\pi_{\vp}}%
\newcommand{\pivpIc}{\pi_{\vpIc}}%
\newcommand{\pivpHIIc}{\pi_{\vpHIIc}}%
\newcommand{\vI}{v_{\I}}%
\newcommand{\trI}{ {\rm{tr}}_{\I}}%
\newcommand{\trori}{ {\rm{tr}}_{\{0\}}}%
\newcommand{\kappaI}{\kappa_{\I}}%
\newcommand{\kappaItil}{\hat{\kappa}_{\I}}%
\newcommand{\piIctil}{\hat{\pi}_{\vpHIIc} }%
\newcommand{\Al}{{\cal{A}}}%
\newcommand{\Ala}{\Al_{\alpha}}%
\newcommand{\Alb}{\Al_{\beta}}%
\newcommand{\Ale}{\Al^{e}}%
\newcommand{\Alo}{\Al^{o}}%
\newcommand{\Alal}{\Al_{\al}}%
\newcommand{\Alalc}{\Al_{\alc}}%
\newcommand{\Alalco}{\Al_{\alc}^{o} }%
\newcommand{\Alae}{\Al_{\alpha}^{e}}%
\newcommand{\Alao}{\Al_{\alpha}^{o} }%
\newcommand{\Albe}{\Al_{\beta}^{e}  }%
\newcommand{\Albo}{\Al_{\beta}^{o}   }%
\newcommand{\Algamma}{\Al_{\gamma}}%
\newcommand{\Algammao}{\Algamma^{o}}%
\newcommand{\Aldel}{\Al_{\del}}%
\newcommand{\Aldelo}{\Aldel^{o}}%
\newcommand{\Aldele}{\Aldel^{e}}%
\newcommand{\Alzeta}{\Al_{\zeta}}%
\newcommand{\Alzetao}{\Alzeta^{o}}%
\newcommand{\AlI}{{\cal{A}}_{\I}}%
\newcommand{\AlIc}{{\cal{A}}_{\Ic}}%
\newcommand{\Alori}{\Al_{\{0\}}}%
\newcommand{\Aloric}{{\cal{A}}_{\oric}}%
\newcommand{\AlIcommut}{\AlI^{\prime}}%
\newcommand{\AlJ}{{\cal{A}}_{\J}}%
\newcommand{\Ali}{\Al_{\{i\}}}%
\newcommand{\AlIe}{\AlI^{e}  }%
\newcommand{\AlIo}{\AlI^{o}   }%
\newcommand{\AlIce}{\AlIc^{e}  }%
\newcommand{\AlIco}{\AlIc^{o}   }%
\newcommand{\Bl}{{\mathfrak B}}%
\newcommand{\ZZ}{{\mathfrak Z}}%
\newcommand{\CENTkappa}{\ZZ_{\kappa}}
\newcommand{\CENTkappae}{\CENTkappa^{e}}   
\newcommand{\CENTkappao}{\CENTkappa^{o}}   
\newcommand{\CENTome}{\ZZ_{\ome}}
\newcommand{\CENTomee}{\CENTome^{e}}
\newcommand{\CENTomeo}{\CENTome^{o}}
\newcommand{\CENTomeehon}{\CENTome^{e({\Thetaome} )}}
\newcommand{\CENTomeohon}{\CENTome^{o({\Thetaome} )}}
\newcommand{\CENTvp}{\ZZ_{\vp}}
\newcommand{\CENTvpIc}{\ZZ_{\vp_{\Ic}}}
\newcommand{\CENTvpHIIc}{\ZZ_{{\vpHI}_{\Ic}}}
\newcommand{\CENTvpHIIce}{\ZZ_{{\vpHI}_{\Ic}}^{e}}
\newcommand{\CENTvpHIIco}{\ZZ_{{\vpHI}_{\Ic}}^{o}}
\newcommand{\Ap}{A_{+}}%
\newcommand{\Am}{A_{-}}%
\newcommand{\Bp}{B_{+}}%
\newcommand{\Bm}{B_{-}}%
\newcommand{\KilI}{{\Kil}_{\I}}%
\newcommand{\Hilome}{{\Hil}_{\ome}}%
\newcommand{\Hilpsitil}{{\Hil}_{\psitil}}%
\newcommand{\Hilvp}{{\Hil}_{\vp}}%
\newcommand{\HilvpHIIc}{\Hil_{{\vpHIIc}}}
\newcommand{\HilvpIc}{\Hil_{{\vpIc}}}
\newcommand{\MPsi}{\mit\Psi}%
\newcommand{\Ome}{\mit\Omega}%
\newcommand{\Omeome}{\Ome_{\ome}}%
\newcommand{\Omepsitil}{\Ome_{\psitil}}%
\newcommand{\Omevp}{\Ome_{\vp}}%
\newcommand{\OmevpIc}{\Ome_{\vpIc}}%
\newcommand{\OmevpHIIc}{\Ome_{{\vpHIIc}}}%
\newcommand{\OmeI}{\Ome_{\I}}%
\newcommand{\Alloc}{\Al_{\rm{loc}}}%
\newcommand{\DB}{D(\Alloc)} %
\newcommand{\potI}{\pot(\I)}%
\newcommand{\potK}{\pot(\K)}%
\newcommand{\potJ}{\pot(\J)}%
\newcommand{\pottJ}{\pott(\J)}%
\newcommand{\HJI}{H_{\J}({\I})}
\newcommand{\HI}{H({\I})}%
\newcommand{\sigvp}{\sigma_{\vp, t}}%
\newcommand{\sigvpHI}{\sigma^{\beta\HI}_{\vp, t} }%
\newcommand{\OmevpHI}{\Ome^{\beta\HI}_{\vp}}
\newcommand{\vNM}{{\mathfrak{M}}}%
\newcommand{\vNN}{{\mathfrak{N}}}%
\newcommand{\vNNf}{\vNN_{1}  }%
\newcommand{\vNNs}{\vNN_{2}  }%
\newcommand{\vNMvp}{\vNM_{\vp}}%
\newcommand{\vNMome}{\vNM_{\ome}}%
\newcommand{\vNMomef}{\vNM_{\omef}}%
\newcommand{\vNMomes}{\vNM_{\omes}}%
\newcommand{\vNMkappa}{\vNM_{\kappa}}%
\newcommand{\vNMkappae}{\vNMkappa^{e}}%
\newcommand{\vNMkappao}{\vNMkappa^{o}}%
\newcommand{\Thetaome}{\Theta_{\ome}}%
\newcommand{\vNNe}{\vNN^{e({\Thetaome} )}}%
\newcommand{\vNNo}{\vNN^{o({\Thetaome})}}%
\newcommand{\vNMomee}{\vNMome^{e}}%
\newcommand{\vNMomeo}{\vNMome^{o}}%
\newcommand{\vNMvpIc}{\vNM_{\vpIc} }%
\newcommand{\vNMvpHIIc}{\vNM_{\vpHIIc} }%
\newcommand{\Sc}{\widetilde{S}}%
\newcommand{\ScI}{\Sc_\I}%
\newcommand{\ScIome}{\ScI(\ome)}%
\newcommand{\FIome}{F_{\I,\beta}^{\pot}(\ome)}
\newcommand{\FIvp}{F_{\I,\beta}^{\pot}(\vp)}
\newcommand{\FIpottpsi}{F_{\I,\beta}^{\pott}(\psi)}
\newcommand{\FIpottpsit}{F_{\I,\beta}^{\pott}(\psi\Theta)}
\newcommand{\FIpottvpHI}{F_{\I,\beta}^{\pott}(\vpHI)}
\newcommand{\potbe}{(\pot,\,\beta)}%
\newcommand{\pottbe}{(\pott,\,\beta)}%
\newcommand{\delal}{\delta_{\al}}%
\newcommand{\Alana}{\Al_{ent}}%
\newcommand{\simT}{\sim^{\Theta}}

\begin{document}
\maketitle
\theoremstyle{plain}%
\newtheorem{lem}{Lemma}
\newtheorem{pro}[lem]{Proposition}%
\newtheorem{df}{Definition}%
\newtheorem{thm}{Theorem}
\begin{abstract}
We discuss fermion grading symmetry 
 for  quasi-local  systems 
with graded commutation relations. 
We introduce 
a   criterion   of 
spontaneously symmetry  breaking  
(SSB)
for  general quasi-local systems.  
It is formulated based on the idea 
 that 
each pair of distinct   phases  
(appeared in spontaneous symmetry breaking)
 should be  disjoint 
not only for   the total system but also  for every  
complementary  
outside system  of a local region  
 specified  by the given   quasi-local  structure.
Under   a completely 
 model independent setting,
we show the  
  absence of  SSB (in the above sense) 
for    fermion grading symmetry.

 We  obtain some  structural results
  for equilibrium states  of  lattice 
systems.
If  there would exist an even   
KMS  state for  some even dynamics 
that is     decomposed   
 into noneven KMS states,
then  those noneven  states    inevitably 
 violate the  local thermal stability condition
 by Araki-Sewell.
\end{abstract}
\section{Introduction}
\label{sec:INTRO}
 The  univalence superselection
 rule  forbids
 the   superposition of two states whose total angular momenta 
 are integers and half-integers.
Surely  the  univalence superselection
is regarded as  a natural law
  \cite{WWW}, see  e.g. 
$\S$ 6.1  of \cite{AOX},  III.1  of \cite{HAAG},
 $\S$ 2.2 of \cite{WEIN}.
This is tantamount  to the 
 unbroken symmetry of the  grading transformations 
that multiply    
fermion fields  by $-1$.
  We clarify that the 
 fermion grading symmetry 
 is  very rigid from a thermodynamic viewpoint 
 and different from other symmetries that can be   broken.

Most readers would consider
 that  our  question is  very   trivial,
  at most only  of mathematical interest,   
saying perhaps  ``There is no indication in  nature 
 to   invalidate this rule. 
Fermions  do not condensate!''
So as  to explain our motivation, 
 let us  recall the correspondence of  fermion
 systems
and  Pauli systems  
(more precisely 
  1D spinless fermion systems
and  spin-1/2 systems)
that are connected by  the Jordan-Wigner transformation 
 \cite{JW}.  
For a finite lattice, 
fermion grading symmetry 
 given as 
\begin{eqnarray*}
\Theta(\ai)=-\ai,\quad  \Theta(\aicr)=-\aicr, 
\ \forall i \in \NN,
\end{eqnarray*}
corresponds   to the 
 Pauli grading 
in the spin-lattice system given as
\begin{eqnarray*}
\theta(\sigma^{x}_{i})=-\sigma^{x}_{i},\ \theta(\sigma^{y}_{i})
=-\sigma^{y}_{i},\ 
\theta(\sigma^{z}_{i})=\sigma^{z}_{i}.
\end{eqnarray*}
For the  infinite chain, however, the 
  correspondence between 
fermion and Pauli systems
 lose  its meaning.
A remarkable thing is that 
 the Pauli grading  can be broken
for some physical models, see  \cite{AMA}
 for the detailed condition of its  symmetry breakdown
for the XY-model.
Such  examples may indicate that  the status of the fermion 
 grading symmetry for  the infinite-dimensional case  
is  not obvious.
We would like to  justify 
 the unbroken symmetry 
 of fermion grading.

We may 
take a    fundamental standpoint for
  superselection rules 
(not  merely  accepting  them)
 considering  that    
 there are  subtle points 
in deciding whether a (conserved) quantity
satisfies the superselection rule \cite{AFOUND}.
There is a  famous theory  
by Doplicher-Haag-Robetrs
 that derives  the 
superselection rules originating from   the  space-time 
 symmetry for 
  Minkowski or  low-dimensional relativistic space-time.
But the  DHR-theory 
 does not cover all    types 
of  superselection rules
appeared  in  nature.
 (The Fermion-Boson alternative and the univalence superselection
 are different things.)

We  are going to   review  some relevant  results on the 
 matehmatical results on the univalence superselection rule as follows. 
If a state is   invariant   under  some  
 asymptotically
 abelian group of automorphisms like  
  space-translations, then   
 fermion grading symmetry is 
  {\it{perfectly}} preserved.
 That is,
any such   state   
has zero  expectation value   for every   
 odd element  \cite{LRO}  \cite{POW}.
(See also  e.g. 7.1.6 of \cite{RU69},
 Exam. 5.2.21 of   \cite{BR}.
 The  same statement 
 for  quantum field theory is given in   \cite{DS}.)
We shall refer to    \cite{NATH}
 that discusses 
  (possible)  forms of 
symmetry breaking of fermion grading transformations for  dynamics
 that commutes   with some   
 asymptotically
 abelian group of automorphisms.  But  
the status of broken and unbroken 
 symmetry of  fermion grading is 
not given there.

It seems not unreasonable to 
expect  unbroken symmetry of  fermion grading    
  irrespective  of such   translation invariant  assumptions.
It has been   shown however that   non-factor quasi-free
states  of the  CAR algebra
 have  odd elements   in their centers and   
 give  an example of the   breakdown of fermion grading symmetry, 
though  being rather technical
 and not coming  from a physical model  
   \cite{MVER}.
We  
note that  two  mutually disjoint  noneven  states
 in the   factor decomposition
of  each   non-factor quasi-free state
  have    a common    state  restriction
  outside of   some   local region.
It can be said that
 those noneven states   are not   macroscopically  distinguishable.

We are  led to consider that the conventional  
criterion  of  
 spontaneously symmetry breaking  based   
  on  
the center  merely for  the total system
is too weak to be an appropriate  formula  for general 
quasi-local systems.
We introduce a more demanding criterion of SSB
 for general quasi-local systems,
 which turns  to be equivalent to
 the usual one for tensor-product systems.
A pair of  states are said to be {\it{disjoint with
 respect to the given quasi-local structure}} 
if for every local region, 
their   state restrictions      
to   its     complementary outside system  
  induce 
 disjoint GNS representations (Definition \ref{df:MDIS}).
Using  this  notion, 
we propose  a criterion of 
  spontaneously symmetry  breaking
(Definition \ref{df:SSB}).

We  show   the absence of 
    spontaneously symmetry  breaking
 in the above sense  for    
   fermion grading symmetry  
for  general graded quasi-local systems
that encompass    lattice and continuous 
 systems (Proposition  
\ref{pro:NOSSB}).
This  proposition 
may be similar  to  
 the following  statement    in 
\cite{ROB}:
No odd element exists 
 in  observable at  infinity \cite{OAI}.

We  study   
 temperature  states
  (Gibbs states and KMS states) 
of   lattice systems with graded commutation relations.
For every  even Gibbs state, 
we have  a grading preserving   isomorphism 
  from  its     center  
 onto  that of  
  its  state restriction to   the   complementary outside  system
 of   each   local region  (Proposition  \ref{pro:HERA}).

 For now, we cannot  provide
 a definite answer whether 
fermion grading symmetry 
is perfectly preserved or not for  temperature states
 of those  lattice systems.
We only claim that
if a  KMS state 
 breaks the fermion grading symmetry, then it is not thermodynamically 
 stable.   
More precisely, suppose that 
 the odd part of the  center of an 
even KMS  state for even dynamics is not empty.
Then  in the factor decomposition of  its  perturbed state 
by a local Hamiltonian  multiplied by 
the inverse temperature,
there are   noneven 
 KMS states 
that violate
the   local thermal stability condition 
(a minimum free energy condition for open systems)
with respect to 
 the  perturbed   dynamics  acting  
trivially on the   specified  local region
(Proposition  \ref{pro:VIOLTS}).
We give a  remark upon  our 
choice of the local thermal stability condition.
In  \cite{AMLTS} 
we introduced two versions
 of   local thermal stability --- LTS-M and LTS-P.
We make use of 
  the latter   that  will  be simply called  
  LTS  here.
 (See  Appendix for  the details.)
Though   we have  no  example  of      
 such breaking   
nor disprove its existence,
  we 
 may say    that the violation of the univalence superselection rule, 
if it would occur,  is  
 pathological  from a thermodynamical viewpoint.
\section{Notation and  some  known  results} 
We recall 
  the definition of quasi-local $\cstar$-systems.
 (For   references, we refer  e.g. to 
 $\S$ 2  of \cite{ROB},  
 $\S$ 2.6 of \cite{BR}, and  $\S$ 7.1 of \cite{RU69}.)
Let $\Fin$ be a directed set with a partial order relation $\ge$
 and an  orthogonal relation $\perp$
 satisfying the following  conditions:\\
\ a) If  
$\al\le\beta$ and $\beta\perp \gamma$, then  $\alpha\perp\gamma$. \\
\ b) For each  $\alpha, \beta\in\Fin$,
 there exists  a   unique upper bound
$\alpha\vee\beta\in \Fin$   
 which satisfies   $\gamma\ge \alpha\vee\beta$ for any $\gamma\in \Fin$  
such that $\gamma\ge \al$
 and  $\gamma\ge \beta$.\\
\ c) For each $\al\in \Fin$, there exists a unique 
 $\alc$ in $\Fin$ 
 satisfying  
$\alc\perp\al$ and
$\alc\ge \beta$ for 
any $\beta\in \Fin$ such that   $\beta\perp\al$.\\

We consider  a $\cstar$-algebra $\Al$ furnished  with 
the following structure.
Let  $\{\Ala; \  \al\in \Fin\}$ be a family of $\cstar$-subalgebras
 of $\Al$ with the index set $\Fin$. 
Let
 $\Theta$ be an   involutive $\ast$-automorphism 
that determines  the  grading on $\Al$ as  
\begin{eqnarray}
\label{eq:EO}
 \Ale := \{A \in \Al \; \bigl| \;   \Theta(A)=A  \},\quad  
 \Alo := \{A \in \Al  \; \bigl| \;  \Theta(A)=-A  \}. 
 \end{eqnarray}
These  $\Ale$ and $\Alo$ are 
called the even and the  odd parts  of $\Al$.
For  $\al\in \Fin$
  \begin{eqnarray}
\label{eq:EO}
 \Alae := \Ale \cap \Alal,\quad
 \Alao := \Alo\cap \Alal.
 \end{eqnarray}
The above  grading structure is
 referred to as fermion grading (by 
 the condition  L4 defined below).
For a given state $\ome$ on $\Al$,
 its restriction to $\Alal$ is denoted $\ome_{\al}$.
If a state  takes zero for    all odd elements,
it is called  even.

Let   $\Finf$ be a subset of $\Fin$ corresponding to  the 
 set of indices of all  local subsystems and 
 set  $\Alloc:=\bigcup_{\al \in\Finf}\Ala$.
We assume L1, L2, L3, L4 as follows:\ \\
\ \\
\ L1. $\Alloc \cap \Aldel$ is norm-dense in $\Aldel$
 for any $\del \in \Fin$.\\
\ L2. If  $\al \ge\beta$, then $\Ala\supset\Alb$.\\
\ L3. $\Theta(\Alal)=\Alal$ for all $\al\in \Fin$.\\
\ L4. For $\al\perp\beta$  the following 
 graded commutation relations
 hold
\begin{eqnarray*}
\label{eq:}
[\Alae,\; \Albe]=0,\quad &&[\Alae,\; \Albo]=[\Alao,\; \Albe]=0,\nonum\\
\{\Alao,\; \Albo\}=0,
\end{eqnarray*}
 where $[A,\ B]=AB-BA$ is the commutator and 
$\{A,\ B\}=AB+BA$ is the anti-commutator.
\ \\

Our $\Finf$ may correspond to   the set of all bounded  open 
 subsets of  a  space(-time) region   
  or the set of all finite subsets 
 of a lattice.
About   the condition c),  $\alc$ will indicate  
the complement  of $\al$
 in the total region.
We set L1  as it is for the necessity in 
  the proof of Proposition 
 \ref{pro:NOSSB}.

For  $A \in \Al$ (and also for $A\in \Alal$ due to 
 the condition L3), we have the 
 following unique decomposition:  
\begin{eqnarray}
\label{eq:EO1}
  A=A_{+} +A_{-},\ \ 
A_{+}:= \frac{1}{2}  \bigl( A + \Theta(A)   \bigr)
\in \Ale (\Alae),\quad 
A_{-}:=\frac{1}{2}  \bigl(A- \Theta(A)   \bigr)\in \Alo (\Alao). 
\end{eqnarray}

In order to ensure that the fermion grading involution 
 $\Theta$
 acts non-trivially on $\Al$,
we may  assume, for example, 
  that     $\Alao$ is not empty 
 for all $\al\in\Fin$.
 However, all our results below
 obviously hold  for any  trivial cases
 where  fermions do not   or  rarely exist.
\section{A criterion of spontaneous symmetry 
 breaking   appropriate for 
general quasi-local systems
and  Fermion grading symmetry}
\label{sec:SSB}
A pair of states 
will be called disjoint with each other 
if their GNS representations are disjoint, see e.g.  $\S$ 2.4.4 
 and $\S$ 4.2.2 of \cite{BR}.
We shall 
employ 
the following  more demanding condition for 
 disjointness of two states.
\begin{df}
\label{df:MDIS}
 Let $\omef$ and $\omes$ be states of a 
quasi-local system $(\Al,\  \{\Ala\}_{\al\in \Finf })$.
If 
 for every  $\gamma\in \Finf$,
their restrictions  
to the complementary outside system of $\gamma$, i.e.,
${\omef}_{{\gammac}}$ and 
${\omes}_{{\gammac}}$
   are disjoint with each other,  
   then $\omef$ and $\omes$ are 
said  to be  disjoint  
 with respect to the quasi-local structure 
$\{\Ala\}_{\al\in \Finf}$.
\end{df}

We shall give  a criterion of 
  spontaneously symmetry breaking  
based on  Definition \ref{df:MDIS}  as follows. 
Let $G$  be   a  group  
and $\tau_{g} (g\in G)$ be its   action of $\ast$-automorphisms
 on a quasi-local system $(\Al,\  \{\Ala\}_{\al\in \Finf })$. 
Suppose that 
$\tau_{g}$ commutes with a  given (Hamiltonian) dynamics for every 
 $g\in G$.
Let $\physical$ denote some set of 
 physical states  (e.g.
the set of all ground states
 or  all  equilibrium states
at some temperature  for the given dynamics), and 
$\physicalG$ denote the set of all $G$-invariant states in $\physical$. 
Let $\ome$ be an extremal point  in  $\physicalG$.
Suppose that   $\ome$ has a factor state decomposition
in $\physical$ in the form of 
  $\ome= \int d\mu(g) \ome_g$ with
$\ome_{g}:=\tau_{g}^{\ast}\ome_{0}
(=\ome_{0}\circ\tau_{g})$,
 where 
$\ome_{0}$  is a factor state 
in $\physical$ (but not in $\physicalG$)
and so is   each   $\ome_{g}$, 
and $\mu$ denotes 
 some  probability
 measure  on $G$.
With the above setting, we define
the following.
\begin{df}
\label{df:SSB}
 If for each  $g\ne g^{\prime}$ of G
a pair of factor states $\ome_{g}$
 and $\ome_{g^{\prime}}$
 are   disjoint with respect to the given quasi-local structure, 
then it is said that 
 the G-symmetry  is  macroscopically  broken.
\end{df}

Let $\ome$ be a state of 
a quasi-local system
 $(\Al,\; \{\Ala\}_{\al\in \Finf })$.
It is said that  $\ome$  satisfies   the cluster property
(with respect to  
 the  quasi-local structure)
 if  for 
  any given $\vareps>0$  and any $A\in \Al$ there exists
 an $\al\in \Finf$ such that
\begin{eqnarray}
\label{eq:CLUS}
\bigl|\ome(AB)-\ome(A)\ome(B)\bigr|<\vareps\Vert B \Vert
\end{eqnarray}
for all $B\in    \bigcup_{\beta\perp\al}\Alb$.
It is  shown in   \cite{ROB}
 and  Theorem 2.6.5 \cite{BR}
 that every factor state  satisfies this  cluster 
 property. However, 
 the converse does not always hold; 
  non-factor quasi-free states 
of the CAR algebra  satisfy the cluster property
 with respect to the 
   quasi-local (lattice) structure used for  their  construction, 
 see  \cite{MVER} for details.

The following   proposition  asserts  that 
 fermion grading symmetry 
cannot be broken   in the sense of 
 Definition \ref{df:SSB}.
A remarkable thing   is that
it  makes no  reference   to the dynamics.
We are using essentially    
no more than  the canonical anticommutation relations (CAR)
 for  its proof.
(The idea of the proof 
comes from   our  study  on     
 state correlation  for  composite  fermion  systems
  done in   \cite{AMEXT} \cite{SEP}.)
\begin{pro}
\label{pro:NOSSB}
Let $\ome$ be a state of a quasi-local system
 $(\Al,\; \{\Ala\}_{\al\in \Finf })$
   and $\Theta$ denote the fermion grading involution of $\Al$.
Suppose that 
$\ome$ satisfies 
the cluster property 
with respect to the quasi-local structure.
Then $\ome$ and $\ome\Theta$
 cannot  be   disjoint  
with respect to the quasi-local structure 
$\{\Ala\}_{\al\in \Finf}$. Accordingly 
 spontaneously  symmetry breaking
in the sense of Definition \ref{df:SSB}
 does not exist 
for fermion grading symmetry. 
\end{pro}
\proof
Suppose  that $\ome$ and $\ome\Theta$
 are disjoint with respect to the quasi-local structure 
$\{\Ala\}_{\al\in \Finf}$.  
Then
 $\ome$ and $\ome\Theta$ restricted to $\Alalc$ are 
 disjoint  for each  $\al\in \Finf$. Hence
 it follows  that
 \begin{eqnarray}
\label{eq:distant}
\bigl\Vert \ome_{{\alc}} - \ome\Theta_{{\alc}}    \bigr\Vert=2.
\end{eqnarray}
This is equivalent to the existence of
 an odd  element  
$\Am\in \Alalco$ such that $\Vert \Am\Vert \le 1$ and  
$| \ome(\Am)- \ome(\Theta(\Am))    |=
| \ome(\Am)- \ome(-\Am)    |=2| \ome(\Am)|=
2$, namely,
 \begin{eqnarray}
\label{eq:distantAm}
 \bigl|\ome(\Am)\bigr|=1.
\end{eqnarray}

By (\ref{eq:EO1})
 and L1, 
  we have that 
 $\Alloc\cap\Aldele$ is norm dense in $\Aldele$
 and   so is $\Alloc\cap\Aldelo$ in $\Aldelo$
 for any $\del\in \Fin$.
Hence from  (\ref{eq:distantAm}), we 
 have some 
 $\Am$ in $\Algammao$
for some 
$\gamma\in \Finf$ 
such that $\gamma\le\alc$, 
$\Vert \Am \Vert\le 1$  and 
 \begin{eqnarray}
\label{eq:distantA}
\bigl|\ome(\Am)\bigr| >0.999.
\end{eqnarray}
(We use    a sloppy    notation for $\Am$
 in the above; $\Am$ in (\ref{eq:distantAm})
 belonging to $\Alalc$
is approximated by $\Am$ in (\ref{eq:distantA})
 belonging to $\Algammao$.)
By  the  decomposition of $\Am$
 into hermitian elements
 \begin{eqnarray*}
\Am=1/2(\Am+\Am^{\ast}) -
i \bigl(i/2 (\Am-\Am^{\ast})\bigr), 
\end{eqnarray*}
we have 
 \begin{eqnarray*}
\bigl| \ome\bigl( 1/2(\Am+\Am^{\ast}) \bigr)-
i \ome \bigl( i/2  (\Am-\Am^{\ast})\bigr)  \bigr|
>0.999.
\end{eqnarray*}
Since  $(\Am+\Am^{\ast})$ and $i (\Am-\Am^{\ast})$
 are both self-adjoint,
 we have
 \begin{eqnarray*}
\bigl|\ome\left(1/2(\Am+\Am^{\ast})\right)\bigr|^{2}
+\bigl|\ome \bigl( i/2 (\Am-\Am^{\ast}) \bigr)\bigr|^{2}  > 0.999^{2}.
\end{eqnarray*}
 Hence  we have
 \begin{eqnarray}
\label{eq:jitsukyo}
\bigl|\ome\left(1/2(\Am+\Am^{\ast})\right)\bigr|
>  \frac{0.999}  {\sqrt{2}}\ \ {\text{or}}\  \ 
\bigl|\ome \bigl( i/2 (\Am-\Am^{\ast}) \bigr)\bigr|
>  \frac{0.999}  {\sqrt{2}}.
\end{eqnarray}
From  (\ref{eq:jitsukyo}),   
$\bigl\Vert 1/2(\Am+\Am^{\ast})\bigr\Vert \le 1$ and  
 $\bigl\Vert i/2 (\Am-\Am^{\ast})\bigr\Vert \le 1$,
we can choose  $\Am=\Am^{\ast}\in \Algammao$ 
(by adjusting $\pm 1$)
such that 
$\Vert \Am \Vert \le 1$ and 
 \begin{eqnarray}
\label{eq:Amsei}
\ome(\Am)  >\frac{0.999}{\sqrt{2}}.
\end{eqnarray}

By   the cluster property assumption  (\ref{eq:CLUS}) on $\ome$,
for a  sufficiently small  $\vareps>0$  and 
the  above specified  $\Am  \in \Algammao$ there exists
 an $\alp\in \Finf$ such that
\begin{eqnarray}
\label{eq:CLUSinproof}
\bigl|\ome(\Am B)-\ome(\Am)\ome(B)\bigr|<\vareps\Vert B \Vert
\end{eqnarray}
for all $B\in    \bigcup_{\beta\perp\alp}\Alb$.

By  (\ref{eq:distant})  with  $\al=\gamma \vee \alp$, 
 the  same argument   leading to (\ref{eq:Amsei}) implies that 
there exists 
 $\Bm=\Bm^{\ast} 
\in  \Alzetao $
such that $\zeta\perp (\gamma \vee \alp)$, $\Vert \Bm \Vert\le 1$
 and 
\begin{eqnarray}
\label{eq:Bmsei}
\ome(\Bm)>\frac{0.999}{\sqrt{2}}.
\end{eqnarray}
Substituting the above $\Bm$
 to  $B$ in (\ref{eq:CLUSinproof}), 
and using (\ref{eq:Amsei}) and  (\ref{eq:Bmsei}), we have
\begin{eqnarray}
\label{eq:yabure}
\bigl| \Imag \left(\ome(\Am \Bm)\right)\bigr|&<&\vareps, \nonum \\
\Real \bigl(\ome(\Am\Bm)\bigr)&>& \frac{0.999^{2}}{2}-\varepsilon.
\end{eqnarray}
 Due to 
$\Am=\Am^{\ast}\in \Algammao$, $\Bm=\Bm^{\ast}\in \Alzetao$, 
 and $\gamma\perp\zeta$,  $\Am\Bm$
 is skew-self-adjoint, i.e.
$(\Am\Bm)^{\ast}=-\Am\Bm$. 
Therefore  
$\ome(\Am\Bm)$
 is a purely imaginary number,
 which however contradicts  with (\ref{eq:yabure}).
Thus we have shown that 
$\ome$ and $\omet$
cannot be   disjoint with respect to  
$\{\Ala\}_{\al\in \Finf}$.  

Since any factor state satisfies the cluster property,
 the   possibility of SSB of Definition \ref{df:SSB}
for the symmetry $\Theta$ is negated. 
\proofend
\section{On the centers of 
temperature  states
of lattice systems}
\label{sec:LATTICE}
From now on, we  consider 
  lattice fermion  systems \cite{AMrmp}
   and also 
the  lattice  systems with graded commutation relations 
\cite{A04}
satisfying  the  translation uniformity   to be specified.
Take   $\Znu$, $\nu(\in \NN)$-dimensional 
cubic integer lattice.
 Let  $\Finf$ be a set  of   all 
 finite subsets  of the lattice.
 We assume that there is  a finite number
of degrees of freedom (spins) on  each site of the lattice.  
For  general graded lattice  systems, we further assume
that 
 the subalgebra $\Ali$ on each site  $i$
 on the lattice
 is   isomorphic to a  $d\times d$ full matrix algebra,
 $d\in \NN$ being independent of $i$.  Hence  
for each  $\I\in \Finf$,  $\AlI$ is isomorphic to
 a  $d^{|\I|}\times d^{|\I|}$ full matrix algebra, 
 and $\Al$ is a UHF algebra of type $d^{\infty}$ 
 by Lemma 2.1 of \cite{A04}.
(As   an example of such  systems, 
$\Ali$
 is generated by 
fermion operators $\ai$, $\aicr$, and 
 spin operators represented by the Pauli matrices 
$\sigma^{x}_{i}$, $\sigma^{y}_{i}$, $\sigma^{z}_{i}$
 which  are even elements    commuting   with all 
fermion operators.)

We denote   the 
 conditional expectation of the tracial state from $\Al$
  onto  $\AlJ$ by $E_{\J}$.
 The interaction among sites  
is determined  by 
the potential $\pot$, a map   from $\Finf$ to $\Al$
 satisfying   the following conditions:\\
\ \\
 $(\pot$-a$)$ $\potI\in \AlI$, $\pot(\emptyset)=0$. \\
  $(\pot$-b$)$  $\potI^{\ast}=\potI$.\\
$(\pot$-c$)$  $\Theta\bigl( \potI \bigr)=\potI$.\\
 $(\pot$-d$)$  $E_{\J} \bigl( \potI\bigr)=0$ 
if $\J \subset \I$ and $\J \ne \I$. \\
$(\pot$-e$)$ For each fixed 
$\I\in \Finf$,
the net $\{\HJI\}_{\J}$ with 
$\HJI:=
\sum_{\K} \bigl\{ \potK;\ \K \cap \I \ne 
\emptyset,\ \K \subset \J \bigr\}$ 
 is a Cauchy net for $\J\in \Finf$ in the norm topology
 converging to a local Hamiltonian 
$\HI\in \Al$.\ \\

Let  $\PB$ denote  the 
 real vector space  of all $\pot$ 
satisfying the  above all conditions.
The set   of all $\ast$-derivations 
on the  domain $\Alloc$ 
 commuting with $\Theta$  is denoted $\DB$. 
There  exists  a bijective real linear map
 from  $\pot\in \PB$ to $\del\in \DB$ 
for the lattice fermion  systems  
(Theorem 5.13  of \cite{AMrmp}), 
and similarly 
   for the graded lattice  systems 
 (Theorem 4.2 of \cite{A04}).
The connection  between  $\del\in \DB$ 
and its corresponding $\pot\in\PB$
 is given by
 \begin{eqnarray}
  \label{eq:DP}
\del(A)=i[\HI,\,A],\quad A \in \AlI 
\end{eqnarray}
 for every $\I\in \Finf$,
  where the  local Hamiltonian  $\HI$ is determined   by 
 $(\pot$-e$)$  for  this    $\pot$.

The condition $(\pot$-d$)$  
is called  the standardness  
  which is for 
fixing   ambiguous terms (such as  scalars) irrelevant to the 
 dynamics given by  (\ref{eq:DP}).
We remark that any   product state, for example 
 the Fock state, can be used
 in place  of the  tracial state for $E_{\J}$
 to obtain a similar  one-to-one correspondence 
 between $\del$ and $\pot$. Furthermore, 
 characterizations  of equilibriums states, 
such as LTS, Gibbs (and also the variational principle
 for  translation  invariant states), have been all 
shown   to be 
independent of  the choice 
 of those product states \cite{A04}.

The above-mentioned  Gibbs condition
   was defined   for the  quantum  spin lattice  systems  \cite{AION},
and then extended to
   the lattice fermion systems  in $\S$ 7.3 
of \cite{AMrmp}, 
and  to  the graded lattice   systems  
under consideration  \cite{A04}.
Let $\Ome$ be a  cyclic and separating vector 
 of a von Neumann algebra $\vNM$ on $\Hil$ and ${\mit\Delta}$
 denote the  modular operator  for $(\vNM, \Ome)$, see
   \cite{TAKE}.
The  state $\ome$ on $\vNM$
 given  by $\ome(A)=(\Ome,A\Ome )$ for $A\in \vNM$
satisfies 
 the KMS condition  for the modular automorphism group
$\sigma_t:= {\mbox {Ad}}(\mit\Delta^{it})$, $t\in \R$, at the 
 inverse temperature $\beta=-1$ and  is  called the 
{\it{modular state}} 
with respect to $\sigma_t$. 
The following definition works
 for   any  lattice  system under  consideration. 
\begin{df}
\label{df:GIB}
Let $\vp$
 be a state of $\Al$ and 
 $\bigl(\Hilvp,\; \pivp,\; \Omevp  \bigr)$
 be   its   GNS triplet.
It is said that $\vp$  satisfies  the  Gibbs condition 
 for $\del\in \DB$ at inverse temperature $\beta\in \R$,
 for short  $(\del,\beta)$-Gibbs condition, 
if  and only if the following conditions are satisfied\;{\rm{:}} \ \\
 {\rm{(}}{\it{Gibbs}}-1\,{\rm{)}} 
The GNS vector  $\Omevp$ is separating for 
 $\vNMvp:=\pi_{\vp}(\Al)^{\prime \prime}$.\ \\
\ For   $(\vNMvp, \Omevp, \Hilvp)$, 
the   modular operator ${\mit\Delta}_{\vp}$ 
 and  the  modular automorphism group 
  $\sigvp$ are defined. 
Let $\sigvpHI$ 
 denote the  one-parameter group of $\ast$-automorphisms
 determined  by the 
 generator $\del_{\vp}+\del_{\pivp(\beta\HI)}$,
 where  $\del_{\vp}$ denotes the generator for $\sigvp$ and  
 $\del_{\pivp(\beta\HI)}(A):=i[\beta\pivp(\HI),\; A]$ for
$A\in \vNMvp$. \ \\
{\rm{(}}{\it{Gibbs}}-2\,{\rm{)}}  
For every $\I\in\Finf$,
$\sigvpHI$
 fixes
 the subalgebra $\pivp(\AlI)$ elementwise.
\end{df}

The modular state  
 for  $\sigvpHI$ 
 is given as   the vector state of  a
 (uniquely determined) unit vector  $\OmevpHI$
 lying in the natural cone for $(\vNMvp, \Omevp)$
 and is denoted   $\vpHI$.
We   use the same symbol  
 for its restriction to  $\Al$, namely,
 $\vpHI(A):=\left(\pi(A)\OmevpHI, 
\OmevpHI\right)$
 for $A\in \Al$.
We remark  that 
$\OmevpHI$ is  normalized 
 and $\vpHI(\id)=1$ in our notation. 
(For the general references of 
the perturbed states, the relative modular automorphisms,
 and their application to quantum statistical mechanics,
 see
\cite{ARArelham} \cite{Aradon} and  $\S$ 5.4 of \cite{BR}.)

We next show the product property of 
 $\vpHI$  in the following sense.
\begin{lem}
\label{lem:GIBprod}
Let $\vp$ be a 
  $(\del,\beta)$-Gibbs state 
for $\del\in \DB$ and $\beta\in \R$. 
If it is even, then  
for each  $\I\in\Finf$,
$\vpHI$ is a   product state 
 extension of the tracial state $\trI$ on  
  $\AlI$ and  its restriction to $\AlIc$, as denoted 
 \begin{eqnarray}
\label{eq:vpHIprod}
\vpHI= \trI\circ \vpHI\bigl|_{\AlIc}. 
\end{eqnarray}
\end{lem}
\proof
It has been  already shown  in Proposition 7.7 
of  \cite{AMrmp} 
for the lattice fermion systems, and
 we can easily verify  this  statement    for 
the graded lattice  systems as well.
 But we shall  provide  a  slightly simpler     proof.

In  Theorem 9.1 of \cite{Aradon} it is shown that
 \begin{eqnarray}
\label{eq:}
\vpHI([Q_1,Q_2] Q)=0
\end{eqnarray}
 for every  $Q_1, Q_2\in \AlI$ and $Q\in \AlIcommut$,
 the commutant of $\AlI$ in $\Al$.
From  this we see that   $\vpHI$ is  a product state extension of 
 the tracial state $\trI$ on $\AlI$
 and its restriction to ${\AlIcommut}$.

Since $\vp$ is an even state and $\HI$ is an even 
self-adjoint element,
 $\vpHI$ is also  even. 
It is easy to see 
\begin{eqnarray*}
\AlIcommut=\AlIce + v_{\I} \,\AlIco,
\end{eqnarray*}
 where 
\begin{eqnarray}
\label{eq:vIEQ}
v_i := \aicr\ai-\ai\aicr,\quad 
 v_{\I} := \prod_{i \in \I}v_i. 
\end{eqnarray}
 This  $v_{\I}$ is a 
self-adjoint unitary  implementing $\Theta$ on $\AlI$.
For $\Ap\in \AlIe$, $\Am \in \AlIo$,
 $\Bp\in \AlIce$ and  $\Bm \in \AlIco$,
 computing  the expectation values of 
all 
$A_{\sigma} B_{\sigma^{\prime}}$ with  $\sigma=\pm$ and  
$\sigma^{\prime}=\pm$
for $\vpHI$,   we    obtain   
 \begin{eqnarray*}
\vpHI(\Ap\Bp)=\trI(\Ap)\vpHI(\Bp),
\end{eqnarray*}
 and  zeros for the others, i.e 
$\Ap\Bm$, $\Am\Bp$ and $\Am\Bm$.
Therefore  $\vpHI$ is 
equal to the 
product state extension of the tracial state $\trI$ on 
 $\AlI$ and $\vpHI|_{\AlIc}$.
\proofend
\ \\

We provide  a grading structure 
 with  von Neumann algebras  generated by   even states
 and  with their $\Theta$-invariant subalgebras.
For an even state $\ome$ of a quasi-local system, 
let 
 $\bigl(\Hilome,\; \piome,\; \Omeome  \bigr)$
be  a GNS triplet of $\ome$
 and let $\vNMome$ denote the von Neumann 
 algebra generated by this representation.
Let $U_{\Theta,\ome}$
be  a unitary operator of $\Hilome$ implementing  
the grading involution $\Theta$,
 and $\Thetaome:
= {\mbox {Ad}}(U_{\Theta,\ome})$.
 Then  even and odd parts of $\vNMome$ are 
  given by
 \begin{eqnarray}
\label{eq:EOvNMome}
 \vNMomee := \{A \in \vNMome \; \bigl| \;   \Theta_{\ome}(A) =A  \},\quad  
 \vNMomeo := \{A \in \vNMome   \; \bigl| \; \Theta_{\ome}(A) =-A  \}. 
 \end{eqnarray}
Let $\vNN$ be a $\Theta$-invariant subalgebra of $\vNMome$.
We give  its grading  as
 \begin{eqnarray}
\label{eq:EOvNN}
 \vNNe  := \vNN \cap \vNMomee,\quad
 \vNNo := \vNN \cap \vNMomeo,
 \end{eqnarray}
where the superscripts $e({\Thetaome} )$
and  $o({\Thetaome} )$  indicate  that 
 the grading is determined by  $\Thetaome$.
For any $A\in \vNMome$ (also  $A \in \vNN$),
 we have its unique decomposition 
  $A=A_{+} +A_{-}$  such that
 $A_{+}\in \vNMomee (\vNNe )$  and 
$A_{-}\in \vNMomeo (\vNNo )$ in the  same manner
  as (\ref{eq:EO1}).

Let $\omef$ and $\omes$ be even states 
 on $\Al$. Let  
 $\vNNf$ and $\vNNs$
be  some $\Theta$-invariant 
subalgebras of $\vNMomef$ and $\vNMomes$, respectively.
If  there is  an isomorphism $\eta$ from
 $\vNNf$ onto    $\vNNs$, that  is,
 $\vNNf$ and   $\vNNs$ are isomorphic, 
 then we denote  this relationship by  $\vNNf\sim\vNNs$.
If  there is  a grading preserving  isomorphism $\eta$ from
 $\vNNf$ onto    $\vNNs$, that  is,
 $\eta$ maps 
 the even part  to the even, the odd to the odd,
 then we write    $\vNNf\simT\vNNs$.
Obviously each of   `$\sim$'
 and `$\simT$' is an  equivalence relation.

We recall 
relative entropy,
which will be used   in the proof of the next
Proposition and also for  the   formulation of our  
local thermal stability condition 
in the next section
 and  Appendix.
For two states  $\omef$  and $\omes$ of a finite-dimensional 
 system, it is  defined by 
 \begin{eqnarray}
\label{eq:REL}
  S(\omef,\ \omes)&:=&
 \omes\left( \log D_2 - \log D_{1}\right),\ {\text{if}}
\ \ker D_{1}\subset \ker D_2,\nonum\\ 
&:=&+\infty,\ {\text{otherwise}},
 \end{eqnarray}
where $D_i$ is the density matrix for $\omei$ $(i=1,2)$.
It is positive, and zero if and only if 
$\omef=\omes$.
Its generalization to     von Neumann
 algebras  is  given in \cite{76REL} \cite{77REL}.
(Note that the order of two states  and the sign  
  convention of relative entropy 
are both reversed in   \cite{BR}.)

In the following discussion we are 
 interested  in  centers.
 Let us denote  the center of 
$\vNMome$
 by $\CENTome$.
 It is immediate  to see that 
$\CENTome$ is $\Theta$-invariant for an even state $\ome$.
We shall use shorthanded 
$\CENTomee$ and  $\CENTomeo$ for   
 $\CENTomeehon$ and $\CENTomeohon$, respectively.

\begin{pro}
\label{pro:HERA}
Let $\vp$ be an even  
  $(\del,\beta)$-Gibbs state. 
For  $\I \in \Finf$, let
 $\vpIc$ denote the state restriction of $\vp$ onto $\AlIc$. 
Then for any  $\I\in \Finf$
there is a grading preserving  isomorphism 
between the centers of the
von Neumann algebras 
generated by the GNS representation 
of $\vp$ and  by that of $\vpIc$. 
Especially, $\vp$
 is a factor state if and only if so is $\vpIc$.
 \end{pro}
\proof
Let 
 $\bigl(\Hilvp,\; \pivp,\; \Omevp  \bigr)$
be  a GNS triplet of $\vp$, and 
    $\OmevpHI$ denote    the normalized  vector
 representing   its  perturbed state $\vpHI$
as  in Definition \ref{df:GIB}.
By Theorem 3.10 of \cite{77REL} 
(also by the discussion below Definition 
 6.2.29 of \cite{BR}),
 \begin{eqnarray}
\label{eq:RELZENTAI}
 S(\vp,\ \vpHI)\le 2 \Vert \beta \HI \Vert,\nonum \\
  S(\vpHI,\ \vp)\le 2 \Vert \beta \HI \Vert.
\end{eqnarray}
Since the relative entropy is not increasing  
by restriction onto any subsystem, 
taking the restrictions of $\vp$ and $\vpHI$
 onto $\AlIc$ denoted $\vpIc$ and $\vpHIIc$ respectively,
we have 
 \begin{eqnarray}
\label{eq:RELIc1}
 S(\vpIc,\ \vpHIIc)\le 2 \Vert \beta \HI \Vert,\\
\label{eq:RELIc2}
  S(\vpHIIc,\ \vpIc)\le 2 \Vert \beta \HI \Vert.
\end{eqnarray}
By applying  the argument in $\S$ 2 and 3 of \cite{AUNI}
 to the present case,
(\ref{eq:RELIc1}) implies that 
$\vpIc$ quasi-contains $\vpHIIc$, and also 
(\ref{eq:RELIc2})  the vice-versa.  
 (The notion of quasi-containment given in this reference is as follows.
 For a pair of representations $\pi_1$ and $\pi_2$
 of a $\cstar$-algebra, if there is a subrepresentation of $\pi_{1}$ 
 which is quasi-equivalent to $\pi_2$, then $\pi_1$ is said to
 quasi-contain $\pi_2$.)
Therefore $\vpIc$ and  $\vpHIIc$ are quasi-equivalent.
Let  
 $\bigl(\HilvpIc,\; \pivpIc,\; \OmevpIc  \bigr)$
and  $\bigl(\HilvpHIIc,\; \pivpHIIc,\; \OmevpHIIc  \bigr)$
be  GNS representations 
 for $\vpIc$ and  $\vpHIIc$,  
$\vNMvpIc$ and  $\vNMvpHIIc$ be  von Neumann algebras 
 generated 
 by those representations of $\AlIc$.
By taking the restriction of 
the canonical isomorphism 
between the von Neumann algebras $\vNMvpIc$ and  $\vNMvpHIIc$
 which maps $\pivpIc(A)$ to $\pivpHIIc(A)$
 for $A\in \Al$ onto their  centers 
$\CENTvpIc:= \vNMvpIc\cap \vNMvpIc^{\prime}$  and 
$\CENTvpHIIc:=\vNMvpHIIc\cap \vNMvpHIIc^{\prime}$,
 we have
 \begin{eqnarray}
\label{eq:ID1}
\CENTvpIc \simT  \CENTvpHIIc.
\end{eqnarray}
In the above derivation, we have noted that  even and odd 
 parts of von Neumann algebras generated by a GNS representation 
are weak limits 
 of  even and odd parts of a 
 underlying $\cstar$-system 
(mapped onto the GNS space), and hence  the canonical 
 isomorphism conjugating   a pair of quasi-equivalent representations  
and its restriction to $\Theta$-invariant subalgebras 
 are  grading preserving.

We  shall construct  a  
 GNS representation  of   $\vpHI$ (on  $\Al$)
 from   the above 
 $\bigl(\HilvpHIIc,\; \pivpHIIc,\; \OmevpHIIc \bigr)$
on $\AlIc$ 
and a GNS representation 
of the tracial state  $\trI$ on $\AlI$ 
denoted 
$\bigl(\KilI, \kappaI, \Ome_{\I}  \bigr)$. 
Define 
\begin{eqnarray}
\label{eq:kappa}
\Kil&:=&  \KilI\otimes \HilvpHIIc,\nonum  \\
\MPsi&:=&  \OmeI  \otimes \OmevpHIIc, 
\nonum \\
V_{\I}&:=& \kappaI(\vI) \otimes \id_{\Ic}, \nonum \\
\kappaItil(A)&:=&   \kappaI(A) \otimes \id_{\Ic}\ {\text{for}} 
\  A\in \AlI,\nonum \\
\piIctil(A)&:=&   \id_{\I} \otimes  \pivpHIIc(A)
\ {\text{for}} 
\  A\in \AlIc,
\end{eqnarray}
 where
$\id_{\I}$ and
$\id_{\Ic}$ are  the identity operators on 
 $\KilI$ and $\HilvpHIIc$, 
$\vI$ is given by  (\ref{eq:vIEQ}).
  Noting   ${\mbox {Ad}}(\vI)=\Theta|_{\AlI}$, 
   we have  a unique representation $\kappa$ 
 of the total  system $\Al$ on  $\Kil$ 
 satisfying
\begin{eqnarray}
\label{eq:kappaI}
\kappa(A)=\kappaItil(A) \ \  {\text{for}}\ \; A\in \AlI, 
\end{eqnarray}
  and 
\begin{eqnarray}
\label{eq:kappaIc}
\kappa(\Bp)=\piIctil(\Bp)\ \  {\text{for}}\ \; \Bp\in \AlIce, \quad
\kappa(\Bm)= V_{\I}\piIctil(\Bm)\ \  {\text{for}}\ \; \Bm\in \AlIco.
\end{eqnarray}
By   (\ref{eq:vpHIprod}), i.e., the product property of $\vpHI$
 for $\AlI$ and $\AlIc$, 
 we verify that 
 this $\bigl(\Kil,\; \kappa,\; \MPsi  \bigr)$
 gives a GNS triplet of $\vpHI$.
We have also 
\begin{eqnarray}
\label{eq:vNMkappa}
\vNMkappa&:=&
\kappa(\Al)^{\prime\prime}= 
\left(\kappaI(\AlI)\otimes 
\pivpHIIc(\AlIc)\right)^{\prime\prime}\nonum\\
&=&\left(\kappaI(\AlI)\right)^{\prime\prime}
\otimes \vNMvpHIIc.
\end{eqnarray}

Since $\vpHIIc$ is even, and  is $\Theta|_{\AlIc}$-invariant,
 we have a unitary operator 
$\UIc$ of $\HilvpHIIc$ 
 which implements $\Theta|_{\AlIc}$ 
 in its GNS space 
 $\bigl(\HilvpHIIc,\; \pivpHIIc,\; \OmevpHIIc  \bigr)$.
As (\ref{eq:EOvNMome}),
${\mbox {Ad}}(\UIc)$ determines
 the   even and odd parts of $\vNMvpHIIc$.
Accordingly by (\ref{eq:EOvNN}),
  the grading is induced on 
the   center  $\CENTvpHIIc$ and it  
 is decomposed  into  $\CENTvpHIIce$ and $\CENTvpHIIco$.

For  $\AlI$,  
$\kappaI(\vI)$  gives   a unitary operator 
  implementing  $\Theta|_{\AlI}$. 
By the construction of 
 $\bigl(\Kil,\; \kappa,\; \MPsi  \bigr)$,
\begin{eqnarray}
\label{eq:Uk}
U:= \kappaI(\vI) \otimes U_{\Ic}
\in \Bl(\Kil)
\end{eqnarray}
  gives   a unitary operator 
which implements  $\Theta$ 
for $\vpHI$.
This $U$ gives   a grading for $\vNMkappa$
 and it is split  into $\vNMkappae$ and $\vNMkappao$.
Also by this grading
the   center  $\CENTkappa:= \vNMkappa\cap \vNMkappa^{\prime}$ 
 is decomposed  into  $\CENTkappae$ and $\CENTkappao$.

  Note  that  the  center of the tensor product of a pair of 
 von Neumann algebras is equal to the 
  tensor product of  their centers by the commutant theorem 
(Corollary 5.11 in I.V. of 
 \cite{TAKE}).
Since $\AlI$ is a full matrix algebra,
 and  the center of any state on it
 is trivial,  by (\ref{eq:vNMkappa}) we have  
\begin{eqnarray}
\label{eq:CENTkappa}
\CENTkappa=
\id_{\I}\otimes  \CENTvpHIIc.
\end{eqnarray}
Moreover from  (\ref{eq:Uk}) and 
(\ref{eq:CENTkappa}) it follows that
\begin{eqnarray}
\label{eq:CENTkappaEO}
\CENTkappa^{e}=
\id_{\I}\otimes  \CENTvpHIIc^{e},\quad 
\CENTkappa^{o}=
\id_{\I}\otimes  \CENTvpHIIc^{o},
\end{eqnarray}
 where we have noted that the grading of  $\CENTvpHIIc$
 is determined  by the unitary  $U_{\Ic}$.
The equalities  (\ref{eq:CENTkappa}) and  (\ref{eq:CENTkappaEO}) 
 give
\begin{eqnarray}
\label{eq:YOWAI}
 \CENTkappa
\simT \CENTvpHIIc.
\end{eqnarray}
Combining (\ref{eq:YOWAI})  
with (\ref{eq:ID1})
we have
\begin{eqnarray}
\label{eq:DESIRE}
\CENTvpIc
\simT \CENTvpHIIc\simT \CENTkappa.
\end{eqnarray}

Since 
  $\bigl(\Kil,\; \kappa,\; \MPsi  \bigr)$
 and  $\bigl(\Hilvp,\; \pivp,\; \OmevpHI  \bigr)$
 are  both GNS representations of the same state 
$\vpHI$ on $\Al$, 
they are apparently unitary equivalent.
The  representation 
$\bigl(\Hilvp,\; \pivp,\; \OmevpHI  \bigr)$
 obviously induces the same von Neumann 
algebra for   $\bigl(\Hilvp,\; \pivp,\; \Omevp  \bigr)$, namely
 $\vNMvp$.
Hence 
  $\bigl(\Kil,\; \kappa,\; \MPsi  \bigr)$
 and  $\bigl(\Hilvp,\; \pivp,\; \Omevp  \bigr)$
 are  unitary equivalent.
Taking the  restriction 
of the unitary map
 which conjugates  those equivalent representations of $\Al$ 
onto the center, we have
\begin{eqnarray}
\label{eq:kappavpEQ}
 \CENTkappa
\simT  \CENTvp.
\end{eqnarray}
From (\ref{eq:DESIRE})
and (\ref{eq:kappavpEQ}), it follows
that
\begin{eqnarray}
\label{eq:assertion}
\CENTvpIc
\simT \CENTvp,
\end{eqnarray}
which is  what we would like to have.
\proofend

\ \\
{\it{Remark 1.}}
We note that  
the  identification 
of  two 
 von Neumann algebras  
in (\ref{eq:YOWAI})
 and  in (\ref{eq:assertion})
does  not imply that  
the underlying $\cstar$-systems $\Al$ and  $\AlIc$
are  conjugated  to each other 
 in those representations.

\ \\
{\it{Remark 2.}}
We   shall explain   that 
the formula
(\ref{eq:assertion})
does not hold  in     general by an example.
Take  one-dimensional lattice $\Z$
 and a site of it, say the origin $0$.
We prepare a 
 non-factor quasi-free state
$\rho$  \cite{MVER} on $\Aloric$, where $\oric$ 
 denote the complementary region  of  $\{0\}$.
The factor decomposition 
of $\rho$ is
given  by  $\rho=1/2(\psi+\psi\Theta)$,
where $\psi$ is a noneven factor state of $\Aloric$.
Take a (unique) product  state extension of 
the tracial state $\trori$ of $\Alori$ 
and $\psi$  to the total system 
$\Al$,
which is denoted $\psitil$.
We see that the state $\psitil\Theta$ on $\Al$ is equal to 
the   state extension of 
$\trori$  and $\psi\Theta$ to $\Al$.
Let 
 $\bigl(\Hilpsitil,\; \pipsitil,\; \Omepsitil  \bigr)$
be  a GNS triplet of $\psitil$.
Take an  odd unitary $u$ of $\Alori$,
 say, $1/\sqrt{2}(a_{0}+a_{0}^{\ast})$.
Define $\xi:=1/\sqrt{2}(\Omepsitil+\pipsitil(u)\Omepsitil) $,
 which is a unit vector of $\Hilpsitil$.
Let   
$\vpx$ denote the  state
determined  by $\vpx(A):=\left(\pipsitil(A)\xi, 
\xi\right)$
 for $A\in \Al$.
It is clear that
this $\vpx$ is a factor state of $\Al$
 by its construction.
By direct computation, 
its restriction onto $\Aloric$ is  
equal to  $\rho$. 
Hence  $\vpx$ is
 a factor state whose restriction to the subsystem
 $\Aloric$ is non-factor. 
\section{Violation of  the local thermal 
 stability  for noneven  KMS 
 states}
\label{sec:VIOLTS}
For some technical reason
 we shall work with KMS states \cite{HHW}
 (not directly with Gibbs states).
Let $\alt$ ($t\in \R$)
 be a one-parameter
group of 
$\ast$-automorphisms  of $\Al$. 
A state $\vp$  is called  an 
  $(\alt,\beta)$-KMS  state 
  if it satisfies 
\begin{eqnarray*}
\vp\bigl(A\al_{i\beta}(B)\bigr)=\vp(BA)
\end{eqnarray*} 
for every
 $A \in \Al$ and $B \in \Alana$, where
 $\Alana$ denotes   the set of all $B \in \Al$
 for which $\alt(B)$ has an analytic extension 
 to $\Al$-valued entire function  $\al_{z}(B)$
 as a function of $z \in \CC$. 

Our  dynamics $\alt$ is assumed to be  even, namely
$\alt\, \Theta=\Theta \,\alt$ for each  $t \in \R$.
We  also put    the following 
 assumptions in order to relate $\alt$ with some $\del \in \DB$. \\
\ (I) The domain  of the generator $\delal$ of $\alt$ 
includes $\Alloc$.\\
\ (II) $\Alloc$ is a core of $\delal$.\ \\

The next  proposition
asserts   the equivalence of the KMS and 
  Gibbs conditions 
under   (I, II).
The proof was  given  
  for the lattice fermion systems in 
 Theorem 7.5 (the implication  
from KMS to Gibbs under the assumption (I)
and Theorem 7.6  (the converse direction  under the assumption 
(I, II))
 of \cite{AMrmp}.
The proof  for  the graded lattice  systems can be 
 done in much the same way  and we shall omit it.
We emphasize that this equivalence 
 does not require the evenness of states,
 which  becomes  essential in  the proof of  Proposition 
\ref{pro:VIOLTS}.
\begin{pro}
\label{pro:KMS-GIB}
Let $\alt$ be an even dynamics satisfying  
the  conditions  ${\rm{(I, II)}}$. 
 Let  $\del(\in \DB)$ be the restriction of its generator $\delal$ 
 to $\Alloc$.
Then a state $\vp$ of $\Al$
 satisfies  $(\alt,\beta)$-KMS condition 
if and only if it 
satisfies  $(\del,\beta)$-Gibbs 
 condition.
\end{pro}

One would ask  whether 
 fermion grading symmetry 
is perfectly preserved or not for non-zero temperature states.
(It is plausible   that  
we can derive   more stronger statement 
 about  the unbroken symmetry of fermion grading 
  for KMS states
than  Proposition \ref{pro:NOSSB}.)
 We leave this question for future study.
 Here we show  the   
 following rather weak   statement.
Suppose that 
there is a nonzero  odd element  in the center of 
 some  even  KMS state for  even dynamics
 satisfying (I, II), 
 then there always exist  noneven KMS states
that do not satisfy the  local thermal stability
 (LTS).
 This LTS  refers to   LTS-P  
in  the terminology of \cite{AMLTS} (not  LTS-M there).
 The  content of the local thermal stability 
 condition is summarized in Appendix.

We shall give  some preparation.
Let $\vp$ be an 
 arbitrary even 
 $(\alt,\beta)$-KMS state.
For    $\I\in \Finf$, which is now fixed,
  $\vpHI$ denotes   the  perturbed state of $\vp$ 
  by $\beta \HI$.
 From the given   $\del\in\DB$ and $\I\in \Finf$,
a new $\ast$-derivation  $\delt\in \DB$  is given  as follows.
Let $\pot\in \PB$ denote  the  potential 
 corresponding to  $\del$.
 Define
 a   new potential $\pott\in \PB$  by    
\begin{eqnarray}
\label{eq:pott}
\pottJ:=0,\ \text{if}\ \J\cap\I\ne\emptyset,\quad 
\text{and}\quad
\pottJ:=\potJ,\ \text{otherwise}.
\end{eqnarray}
We denote 
 the $\ast$-derivation 
corresponding to $\pott$
 by  
 $\delt \in \DB$.
By definition, $\delt$ acts trivially on $\AlI$.
The 
  one-parameter
group of $\ast$-automorphisms of $\Al$
  generated by  $\delt$ is equal to  
 the 
 perturbation  of $\alt$
 by $\HI$ given in  terms of the   
Dyson-Schwinger expansion series and denoted $\altt$.  
By   
Proposition 
\ref{pro:KMS-GIB} and   its proof  
found   in  \cite{AMrmp},
$\vpHI$ satisfies 
$(\altt,\beta)$-KMS  condition and 
 $(\delt,\beta)$-Gibbs condition.

We recall   the  GNS representation 
 $\bigl(\Kil,\; \kappa,\; \MPsi  \bigr)$
of $\vpHI$ previously defined 
 in  (\ref{eq:kappa})  (\ref{eq:kappaI})  (\ref{eq:kappaIc}).
Let 
$p$ be a  nonzero  projection  in
$\CENTkappa$  which has a
 unique even-odd decomposition 
 $p=p_{+} +p_{-}$,   $p_{+}\in \CENTkappae$
 and  $p_{-}\in \CENTkappao$.
By (\ref{eq:CENTkappa}) we can write 
$p=\id_{\I}\otimes q$  with some $q \in \CENTvpHIIc$.
Furthermore by  
(\ref{eq:CENTkappaEO}),   
we have  $p_{+}=\id_{\I}\otimes \qp$ with 
$\qp\in  \CENTvpHIIc^{e}$ and 
$p_{-}=\id_{\I}\otimes \qm$ with 
$\qm\in  \CENTvpHIIc^{o}$.
We  define a positive linear functional on $\Al$ by
\begin{eqnarray}
\label{eq:functional}
\vpHIp(A):=\left(\kappa(A)  \MPsi, 
p \MPsi \right)\ \ {\text{for}}\ A\in\Al.
\end{eqnarray}
We take its restriction onto $\AlIc$.
For $\Ap \in \AlIce$, 
we have 
\begin{eqnarray}
\label{eq:KEEPEV}
\vpHIp(\Ap)&=&\left(\kappa(\Ap)\MPsi, 
p \MPsi\right)\nonum \\
&=&\left(  \left(
 \id_{\I} \otimes  \pivpHIIc\!\!\!(\Ap) \right)
\,\OmeI  \otimes \OmevpHIIc,\, 
 \OmeI  \otimes q \OmevpHIIc \right) \nonum \\
&=&\left( \pivpHIIc\!\!\!(\Ap)
 \OmevpHIIc,\, 
 q \OmevpHIIc \right)\nonum\\
&=&\left( \pivpHIIc\!\!\!(\Ap)
 \OmevpHIIc,\, 
 \qp \OmevpHIIc \right),
\end{eqnarray}
where in the last equality we have used 
 the evenness of $\vpHI$.
For  $\Am\in \AlIco$, 
\begin{eqnarray}
\label{eq:kieru}
\vpHIp(\Am)&=&\left(\kappa(\Am)\MPsi, 
p \MPsi\right)\nonum \\
&=&\left(  \left(
 \kappaI(\vI) \otimes  \pivpHIIc(\Am) \right)
\OmeI  \otimes \OmevpHIIc, 
 \OmeI  \otimes q \OmevpHIIc \right) \nonum \\
&=&
\trI(\vI)
\left( \pivpHIIc(\Am)
 \OmevpHIIc, 
 q \OmevpHIIc \right)=0,
\end{eqnarray}
where we have   used   $\trI(\vI)=0$.

If  $p$ is even, i.e. 
$p=p_{+}=\id_{\I}\otimes \qp$ with 
$\qp \in  \CENTvpHIIc^{e}$,
then from (\ref{eq:KEEPEV}), (\ref{eq:kieru}),
 and $\vpHI(\Am)=0$ for any $\Am\in \Alo$, 
it follows that 
\begin{eqnarray}
\label{eq:evenp}
\vpHIp(A)=
\left( \pivpHIIc(A)
 \OmevpHIIc, 
 \qp \OmevpHIIc \right)
\end{eqnarray}
for any $A\in\AlIc$.
%
%

Suppose that  $\CENTkappao$ is not  empty.
 Take  any nonzero $f\in \CENTkappao$.
Then  $f+f^{\ast}$
 and 
$if+(if)^{\ast}$
 are self-adjoint elements in $\CENTkappao$.
Since at least one of them is nonzero,
   we can take  
a self-adjoint element  in  $\CENTkappao$ 
 whose operator norm is less than $1$ and shall
 denote such  element  by   $f$.
Let  $\pf:=1/2(1+f)$,
which   is a positive operator. 
 Define a noneven state   
\begin{eqnarray}
\label{eq:PSI}
\psi:=2 \vpHIpf
\end{eqnarray}
 by substituting this $\pf$
 into $p$ of (\ref{eq:functional}).
We easily see that  $\psi\Theta$ is equal to $2 \vpHIpfm$
 for $\pfm:=1/2(1-f)$.
 Their averaged state   $1/2(\psi+\psi\Theta)$
 is obviously equal to $\vpHI$.
\begin{pro}
\label{pro:VIOLTS}
Let  $\alt$
be an even dynamics  satisfying 
 {\rm{(I, II)}} and let
 $\vp$ be an arbitrary even 
 $(\alt,\beta)$-KMS state.
 For
 $\I\in\Finf$,
let $\altt$
denote   the  perturbed dynamics
of $\alt$ by 
 the local Hamiltonian $\HI$.
Let $\pott$ denote  the potential for $\altt$ 
given  as (\ref{eq:pott}).
If the odd part of  the center of 
the perturbed state $\vpHI$ is not empty, 
then the  noneven 
 $(\altt,\beta)$-KMS states
$\psi$ and $\psi\Theta$
 given as (\ref{eq:PSI}) 
 violate  $\pottbe$-LTS
 condition.
 \end{pro}
\proof
Since  $\vpHI$
is an  
$(\altt,\beta)$-KMS state,
  $\psi$  and $\psi\Theta$
 are  also    $(\altt,\beta)$-KMS states
by  Theorem 5.3.30 \cite{BR}.
 Accordingly
$\vpHI$,  $\psi$  and $\psi\Theta$ 
 are 
all $(\delt,\beta)$-Gibbs states by Proposition \ref{pro:KMS-GIB}. 

We consider the state restrictions
of  $\vpHI$, $\psi$, and $\psi\Theta$
 onto $\AlIc$.
Since  the even parts of
 $\pf$ and $\pfm$ are  both scalar,
 it follows from   
(\ref{eq:KEEPEV})
that 
\begin{eqnarray*}
\label{eq:RESTIc}
\vpHI|_{\AlIce}=\psi|_{\AlIce}=
\psi\Theta|_{\AlIce}.
\end{eqnarray*}
Due to  to (\ref{eq:kieru})  all of them 
are even when restricted to $\AlIc$. Hence    
we have 
\begin{eqnarray}
\label{eq:RESTIc}
\vpHI|_{\AlIc}=\psi|_{\AlIc}=
\psi\Theta|_{\AlIc}.
\end{eqnarray}

Denote the local Hamiltonians for the new potential 
$\pott$ determined by 
the formula $(\pot$-e$)$ 
by  $\{\HJt\}_{\J\in\Finf}$.
From  (\ref{eq:pott}) it follows that
\begin{eqnarray*}
\label{eq:}
\HIt=0,
\end{eqnarray*}
and hence
\begin{eqnarray}
\label{eq:HIzero}
\vpHI(\HIt)=\psi(\HIt)=\psi\Theta(\HIt)=0.
\end{eqnarray}

 We compute  conditional entropy  
of $\vpHI$, $\psi$ and $\psi\Theta$
for the finite region   $\I$.
The definition of conditional entropy 
 is given  in  (\ref{eq:SCIDEF}). 
Noting    (\ref{eq:vpHIprod}) we have  
\begin{eqnarray}
\label{eq:ScIvpHI}
\ScI(\vpHI)=-S(\trI\circ \vpHI|_{\AlIc},\,\vpHI)=
-S(\trI\circ \vpHI|_{\AlIc},\,\trI\circ \vpHI|_{\AlIc})=0,
\end{eqnarray}
which is the maximum value of $\ScI(\cdot)$.

For  $\psi$,
using 
(\ref{eq:RESTIc}) and then  (\ref{eq:vpHIprod})  we have
\begin{eqnarray}
\label{eq:}
\ScI(\psi)&=&-S(\trI\circ \psi|_{\AlIc},\,\psi)\nonum\\
&=&-S(\trI\circ \vpHI|_{\AlIc},\,\psi)\nonum\\
&=&-S(\vpHI,\,\psi).
\end{eqnarray}
Since $\vpHI\ne\psi$, the former is even and 
the latter is noneven, 
it follows from this equality and  the 
 strictly positivity of relative entropy 
(see \cite{76REL}) that 
\begin{eqnarray*}
\ScI(\psi)<0.
\end{eqnarray*}
By the automorphism invariance 
(acting two states in the argument)
 of relative entropy,
we have 
\begin{eqnarray}
\label{eq:ScImin}
\ScI(\psi\Theta)=\ScI(\psi)<0.
\end{eqnarray}

Substituting  (\ref{eq:HIzero}),
 (\ref{eq:ScIvpHI}), and 
(\ref{eq:ScImin})
into (\ref{eq:FIM}),
we obtain
\begin{eqnarray}
\label{eq:violate}
\FIpottpsi=\FIpottpsit<\FIpottvpHI=0. 
\end{eqnarray}
This strict inequality 
 with  (\ref{eq:RESTIc}) shows that 
$\psi$ and  $\psi\Theta$
 do  not satisfy  
 $\pottbe$-LTS condition (\ref{eq:dfLTSM}),
 although  both of them  satisfy   $(\delt,\beta)$-Gibbs condition. 
\proofend
{}
\newpage
\appendix
\section{Appendix}
\label{sec:APP}
\subsection{Local thermal stability (LTS) condition} 
 Let  $(\Al,\; \{\AlI\}_{\I \in \Finf })$
 be  a  lattice system
  considered  in $\S$ \ref{sec:LATTICE}.
In \cite{AMLTS} the  local thermal stability (LTS) is   
 studied   for the lattice  fermion systems.
 It is easy to see that the same  formulation 
  is available  for  
 the graded lattice   systems under consideration.

 Let $\ome$ be a state of   $(\Al,\; \{\AlI\}_{\I \in \Finf })$.
For $\I\in \Finf$, the conditional entropy of $\ome$ is 
defined  in terms of 
  the relative entropy (\ref{eq:REL}) by 
\begin{eqnarray}
\label{eq:SCIDEF}
\ScIome:=-S(\trI\circ \ome|_{\AlIc},\,\ome)=
-S(\ome\cdot E_{\Ic},\,\ome)\le 0,
\end{eqnarray}
where $E_{\Ic}$ is the conditional expectation 
 onto $\AlIc$  with respect to the tracial state
 and $\ome\cdot E_{\Ic}(A):=\ome( E_{\Ic}(A))$ for $A\in \Al$.

Let $\pot\in \PB$.
 The conditional free energy  
of $\ome$ for $\I\in \Finf$ 
is given by
\begin{eqnarray}
\label{eq:FIM}
\FIome:=\ScIome -\beta\ome(\HI), 
\end{eqnarray}
 where $\HI$ is a local Hamiltonian  for  
 $\I$ with respect to $\pot$.
\begin{df}
\label{df:LTS}
Let  $\pot$ be a  potential in $\PB$.
 A state $\vp$ of $\Al$ is said to satisfy 
the local thermal stability condition 
 for $\pot$ at inverse temperature $\beta$ or
 $\potbe$-LTS
 condition  if for each 
  $\I\in \Finf$
\begin{eqnarray}
\label{eq:dfLTSM}
\FIvp\ge\FIome
\end{eqnarray}
for any state $\ome$ satisfying $\ome|_{\AlIc}=\vp|_{\AlIc}$.
\end{df}
There is  the other definition of    local thermal stability 
in \cite{AMLTS} 
that has the same  variational 
 principle  formula as above but  
takes   the commutant algebra  $\AlI^{\prime}$
 as  the complementary outside system of a local region $\I$ 
instead of $\AlIc$.
 We shall call this alternative 
local thermal stability condition   LTS$^{\prime}$ condition, where the 
 superscript   `$\prime$'
 stands for  the commutant.  (Also by  `$\prime$'
 we mean  that this formalism   is not so  natural
 compared to    Definition \ref{df:LTS}
 if  we  respect   the given  quasi-local structure.
 Nevertheless,   there are  some 
{\it{mathematically}} 
good points with   LTS$^{\prime}$  as will be noted 
in the next paragraph.)

The equivalence of KMS and  LTS$^{\prime}$ conditions 
 holds  for the lattice fermion  systems 
 without assuming the evenness on  states.
 For our  LTS, on the contrary,  
 such  evenness assumption
is  required in deriving  its   equivalence  to the 
 KMS  condition. 
(The  formalism of  LTS$^{\prime}$
 using    commutants  
for complementary outside systems 
  makes it possible to  exploit  the  known 
arguments     
   for   quantum  spin lattice   systems   \cite{ASEW}.)
\end{document}